\documentclass[10pt]{article}

\usepackage[letterpaper, textwidth=5.5in, textheight=9in, top=1in, headsep=0pt]{geometry}
\usepackage[numbers, sort&compress]{natbib}
\usepackage[utf8]{inputenc} 
\usepackage[T1]{fontenc}    
\usepackage{times}          
\usepackage{hyperref}       
\usepackage{url}            
\usepackage{booktabs}       
\usepackage{amsfonts}       
\usepackage{nicefrac}       
\usepackage{microtype}      
\usepackage{xcolor}         
\usepackage{tabularx}       
\usepackage{array}
\usepackage{enumitem}
\usepackage{titlesec}

\hypersetup{colorlinks=true, linkcolor=black, citecolor=blue!60!black, urlcolor=blue!60!black}

\titleformat{\section}{\large\bfseries}{\thesection}{1em}{}
\titleformat{\subsection}{\normalsize\bfseries}{\thesubsection}{1em}{}
\titleformat{\subsubsection}{\normalsize\bfseries}{\thesubsubsection}{1em}{}
\titlespacing*{\section}{0pt}{12pt}{6pt}
\titlespacing*{\subsection}{0pt}{9pt}{4pt}
\titlespacing*{\subsubsection}{0pt}{8pt}{3pt}
\titlespacing*{\paragraph}{0pt}{5pt}{1em}

\newenvironment{ack}{\section*{Acknowledgments}}{}

\newcolumntype{L}{>{\raggedright\arraybackslash}X}

\begin{document}

\begin{center}
  \rule{\linewidth}{4pt}\par\vspace{0.25in}
  {\Large\bfseries The Watchtower Imperative: Lifespan-Calibrated\\[2pt] Evaluation for Frontier AI\par}
  \vspace{0.22in}\rule{\linewidth}{1pt}\par
  \vspace{0.3in}
  \begin{tabular}[t]{c}
    \textbf{Casey M.~Williams}\\
    University of Kansas\\
    \texttt{casey.williams@ku.edu}
  \end{tabular}\hspace{3em}%
  \begin{tabular}[t]{c}
    \textbf{Jennifer Victoria Scurrell}$^{*}$\\
    University of Zurich\\
    \texttt{jennifer.scurrell@ipz.uzh.ch}
  \end{tabular}
\end{center}
\renewcommand{\thefootnote}{\fnsymbol{footnote}}
\footnotetext[1]{Corresponding author.}
\renewcommand{\thefootnote}{\arabic{footnote}}
\vspace{0.2in}

\begin{center}{\large\bfseries Abstract}\end{center}
\vspace{-6pt}
\begin{quote}
  The gap between AI capability deployment and safety assurance is structural and cannot be closed through internal evaluations alone. Analyzing documented harms (2017--2026) across the human lifespan reveals that disparate incidents --- from voice assistants endangering children to AI companions implicated in adolescent suicide and automated care denials for the elderly --- reflect a single failure: evaluating systems against an implicit adult user model. We introduce five developmental moderators structured into a Lifespan Vulnerability Matrix across six life stages. The matrix reveals a non-monotonic risk distribution: vulnerability peaks at both ends of the lifespan, governance coverage is asymmetrical, and authority to convert model errors into real-world impact is scrutinized least. We then propose a set of lifespan design principles generalizing existing child-protective instruments, and an institution-agnostic, three-tier external evaluation framework --- comprising automated simulation, expert red teaming, and post-deployment monitoring --- indexed directly to matrix risk cells.
\end{quote}

\noindent\textbf{Keywords:} AI safety, external red teaming, continuous monitoring, structured escalation, developmental psychology, lifespan development, cognitive aging, AI governance

\medskip
\noindent\textit{Content note: this paper discusses suicide, child safety, CSAM, and elder exploitation in the context of documented AI safety incidents.}

\section{Introduction}

In December 2021, a voice assistant instructed a ten-year-old to touch a coin to a live plug \citep{bbc2021alexa}. In February 2024, a fourteen-year-old died by suicide after an AI companion cultivated dependency and ignored suicidal ideation \citep{garcia2024}. In November 2023, an insurer was alleged to have used predictive algorithms to systematically deny post-acute care to elderly enrollees \citep{lokken2023, ross2023}. In February 2024, a finance employee transferred \$25 million following a deepfake video conference with synthetic colleagues \citep{cnn2024deepfake}.

Though assigned to disparate literatures --- child safety, adolescent mental health, healthcare accountability, and enterprise fraud --- these incidents stem from a single structural error: general-purpose AI systems are evaluated against an implicit model of a cognitively typical adult and deployed to a population where that user is a minority. What varies is not system failure, but human developmental state.

Consequently, we argue that independent external AI safety evaluation must be organized around developmental state, which determines how system behaviors convert into injury. Even a rigorous, adversarial evaluation regime will miss most harms if calibrated solely to a modal adult.

Recent events highlight the urgency of this gap. In mid-2026, frontier models demonstrated autonomous capability to identify large numbers of zero-day vulnerabilities \citep{glasswing, mythos}. The rapid deployment, suspension, and redeployment cycle that followed was governed by capability disclosures and trade policy \citep{fablestatement}, without independent evaluation bodies positioned to contest deployment decisions.

This paper proceeds as follows: Section~\ref{sec:record} surveys the empirical record of AI harms (2017--2026) by developmental stage. Section~\ref{sec:basis} details the Lifespan Vulnerability Matrix, its five moderators, and a proposed set of lifespan design principles. Section~\ref{sec:internal} diagnoses why internal evaluations fail and examines cross-industry precedents. Section~\ref{sec:framework} presents a three-tier external governance framework. Sections~\ref{sec:governance}--\ref{sec:conclusion} outline governance recommendations, address limitations, and conclude.

\section{The Empirical Record}
\label{sec:record}

This section reviews the documented evidence of AI safety failures, tracing both the quantitative surge in reported incidents and their real-world impact. While safety analyses conventionally categorize failures by model behavior, evaluating harm requires examining human vulnerability. We first present macro-level metrics on incident growth before reorganizing the empirical record across human developmental stages.

\subsection{The Acceleration of Documented Harm}

Documented AI harm incidents have more than doubled since 2023. The Stanford AI Index recorded a 56.4\% year-over-year increase for 2023--2024 (149 to 233 incidents) \citep{hai2025}, followed by a further rise to 362 incidents in 2025 \citep{hai2026}. Over the same period CrowdStrike observed an 89\% year-over-year increase in attacks conducted by AI-enabled adversaries \citep{crowdstrike2026}.

These metrics represent severe underestimates due to commercial disincentives and high statutory reporting thresholds (e.g., the EU AI Act \citep{euaiact}; California SB-53 \citep{sb53}). Rather than total volume, rising incident curves signal a widening gap between deployment pace and assurance capacity. Table~\ref{tab:indicators} summarizes key indicators.

\begin{table}[t]
  \caption{Indicators of the AI safety monitoring gap.}
  \label{tab:indicators}
  \centering
  \small
  \begin{tabularx}{\linewidth}{@{}L>{\raggedright\arraybackslash}p{2.6cm}>{\raggedright\arraybackslash}p{4.0cm}@{}}
    \toprule
    Metric & Finding & Source \\
    \midrule
    Incident growth 2023$\rightarrow$2024 & +56.4\% YOY & Stanford AI Index 2025 \citep{hai2025} \\
    Incident growth 2024$\rightarrow$2025 & 233 $\rightarrow$ 362 & Stanford AI Index 2026 \citep{hai2026} \\
    Attacks by AI-enabled adversaries & +89\% YOY & CrowdStrike 2026 \citep{crowdstrike2026} \\
    Vulnerabilities exploited before public disclosure & 42\% & CrowdStrike 2026 \citep{crowdstrike2026} \\
    FDA-registered AI health tools with a post-deployment surveillance plan & 9\% & Dolin et al.\ (2026) \citep{dolin2026} \\
    Organizations actively monitoring production AI & $<$50\% (9\% of small firms) & Gradient Flow (2025) \citep{gradientflow2025} \\
    Teens (13--17) who have used AI companions & 72\% & Common Sense Media (2025) \citep{commonsense2025} \\
    Companion users spending at least as much time with AI as with friends & $\approx$20\%\textsuperscript{a} & Common Sense Media (2025) \citep{commonsense2025} \\
    Adolescent depression per daily social media hour & +13\% odds & Liu et al.\ (2022) \citep{liu2022} \\
    US ad revenue from users under 18, major social platforms (2022) & \$11 billion & Raffoul et al.\ (2023) \citep{raffoul2023} \\
    \bottomrule
  \end{tabularx}

  \smallskip
  \begin{minipage}{\linewidth}
  \footnotesize\textsuperscript{a} Derived: 80\% of companion users report spending more time with friends and 6\% report the reverse; the residual reports parity. Denominator is companion users, not all teens.
  \end{minipage}
\end{table}

\subsection{Harm Incidents by Developmental Stage}

Conventional taxonomies categorize harms by failure mechanism (e.g., hallucination, bias). We reorganize incidents by human developmental stage, which dictates severity and risk exposure.

\subsubsection{Early and Middle Childhood (0--12)}

Children in this range cannot reliably evaluate conversational AI as a fallible, commercially motivated artifact. First-order theory-of-mind consolidates by roughly age five \citep{wellman2001}, but the capacity to model institutional and commercial intent behind a persuasive interface develops considerably later; existing child-protection instruments accordingly treat age-appropriate design as a producer obligation rather than a comprehension assumption \citep{ieee2089, ico2020, crc2021}. Documented incidents include voice assistants instructing children to perform dangerous acts \citep{bbc2021alexa}, insecure smart toys permitting unauthorized surveillance and communication \citep{bnetza2017}, and chatbots lacking age-verification safeguards \citep{garante2023}. These failures are consistently detected reactively, after internal developer reviews fail to account for child developmental limitations.

\subsubsection{Adolescence (13--17)}

Adolescents face an asymmetry between peak reward sensitivity and maturing regulatory control \citep{orben2022, steinberg2008}. Engagement-optimized companion AIs have been alleged to cultivate dependency, discourage help-seeking, and supply suicidal ideation and method details, leading to fatalities and litigation \citep{cnn2025chatgpt, garcia2024}. Further harms include chatbots advising minors on concealing alcohol and sexual activity \citep{ccdh2023}, an eating-disorder support chatbot offering harmful weight-loss guidance \citep{npr2023}, and recommendation algorithms rapidly delivering self-harm content \citep{ccdh2022, russell2022}.

\subsubsection{Emerging Adulthood (18--25)}

Despite ongoing prefrontal maturation and skill formation, regulatory oversight terminates at age 18. Documented risks include chatbot-mediated radicalization preceding violent crime \citep{bbc2023windsor} and false academic integrity accusations from detection tools \citep{wapo2023}. A third risk --- erosion of professional judgment through premature reliance on automated outputs during early-career skill acquisition --- follows from the cognitive offloading literature \citep{risko2016} but has not been systematically documented in this population, a gap we return to in Section~\ref{sec:limitations}.

\subsubsection{Established Adulthood (26--64)}

Adults possess authority to translate model errors into real-world consequences, exacerbated by automation bias \citep{parasuraman2010} and Bainbridge's irony \citep{bainbridge1983}. Documented failures include legal briefs citing hallucinated precedents \citep{gauthier2024, mata2023}, binding corporate liability for chatbot misrepresentations \citep{moffatt2024}, false criminal allegations generated about a named individual \citep{noyb2025}, wrongful arrests via facial recognition errors \citep{williams2020, woodruff2023}, systemic administrative fraud misdeterminations \citep{robodebt2023}, batch adjudication of insurance claims without individual review \citep{propublica2023}, and large-scale financial fraud via deepfake video calls \citep{cnn2024deepfake}.

\subsubsection{Later Adulthood (65+)}

Older adults experience cognitive decline \citep{salthouse2009} and elevated risk of social isolation \citep{holtlunstad2015} while retaining financial and medical authority. Principal documented harms involve alleged algorithmic denial of post-acute care via predictive insurance software \citep{lokken2023, ross2023, psi2024} and AI-enabled voice cloning and extortion fraud \citep{fbi2024, senate2023voice}, which exploit susceptibility patterns established in the pre-AI elder-fraud literature \citep{deliema2018}. Conversational companion robotics have been deployed at state scale as a social isolation intervention \citep{nysofa2022}; whether such systems supplement or displace human contact is, in our view, an open empirical question that current deployment evaluations are not designed to answer.

\subsubsection{Stage-Independent Harm Domains}

Three harm domains affect all populations:
\begin{enumerate}[label=(\arabic*), leftmargin=*, itemsep=2pt]
  \item \textbf{Physical Safety:} Navigation and autonomous driving failures causing fatalities and traffic disruption \citep{aiid, oecd2025}.
  \item \textbf{Adversarial Exploitation:} Systematic prompt vulnerabilities, red-teaming failures, and automated espionage campaigns \citep{sabin2025, techradar2025, ukaisi2024}.
  \item \textbf{Democratic Integrity:} Deepfakes and AI disinformation targeting elections \citep{aiid, oecd2025}.
\end{enumerate}
Systemic market concentration \citep{bengio2025, bengio2026} amplifies these risks: model miscalibrations propagate simultaneously across all developmental stages.

\section{The Developmental Basis}
\label{sec:basis}

This section provides the theoretical foundation for developmental AI safety evaluation. Rather than viewing harm through isolated system failures, we establish a framework centered on human cognitive and social vulnerability. We first define five stage-varying moderators, map them across six lifespan stages in a Lifespan Vulnerability Matrix, and then propose a set of lifespan design principles generalizing existing child-protection instruments.

\subsection{Five Developmental Moderators}

Converting the empirical record into an evaluation specification requires constructs that are stage-independent in definition but stage-varying in value. We propose five moderators that determine how system behaviors convert into injury and how severely:

\paragraph{M1. Attribution capacity.} The user's ability to accurately model the system's nature, limits, incentives, and epistemic status. Absent before theory-of-mind consolidates \citep{wellman2001}, near ceiling in healthy adulthood, and declining with cognitive aging. Low attribution converts fluency into misplaced trust. Candidate measures: calibration of user confidence against system accuracy; elicited beliefs about agency and memory.

\paragraph{M2. Self-regulatory capacity.} The ability to disengage from reinforcing interactions. Suppressed in adolescence by reward--control asymmetry \citep{steinberg2008}, highest in established adulthood, and declining in later life alongside broader executive change \citep{salthouse2009}. Low capacity converts engagement optimization into compulsive use. Candidate measures: session lengths, disengagement latency, return frequency.

\paragraph{M3. Skill-formation exposure.} Whether an offloaded capability is under active acquisition, consolidated, or under maintenance. Offloading consolidated skills improves efficiency. We hypothesize that offloading skills under active formation produces developmental displacement: the offloading literature establishes the underlying mechanism in adult populations \citep{risko2016, sparrow2011}, but has not been extended to users who are still acquiring the offloaded capability. We treat this as a measurement priority rather than an established finding (Section~\ref{sec:future}). Relevant during foundational learning, early-career skill acquisition, and cognitive maintenance in aging populations.

\paragraph{M4. Consequence authority.} The real-world stakes a user can commit through an AI-mediated decision. Minimal in childhood, peaking in established adulthood, and remaining high for financial and medical decisions in later life. Consequence authority acts as a multiplier on factual error: identical hallucination rates produce vastly different harms depending on user authorization scope.

\paragraph{M5. Social substitution risk.} The likelihood that AI interactions displace human relationships rather than supplementing them. U-shaped across the lifespan: elevated in adolescence during identity formation \citep{commonsense2025, orben2022} and in later adulthood due to social network contraction \citep{carstensen2006, holtlunstad2015}. Candidate measures: longitudinal AI-to-human interaction ratios; displacement of reported contacts.

\subsection{The Lifespan Vulnerability Matrix}

Table~\ref{tab:matrix} maps these five moderators across six developmental stages, linking elevated moderators to dominant harm vectors and evaluation requirements.

\begin{table}[t]
  \caption{The Lifespan Vulnerability Matrix. Listed moderators are those elevated at that stage; unlisted moderators are present but not governing.}
  \label{tab:matrix}
  \centering
  \small
  \setlength{\tabcolsep}{4pt}
  \begin{tabularx}{\linewidth}{@{}>{\raggedright\arraybackslash}p{2.1cm}>{\raggedright\arraybackslash}p{1.5cm}LL@{}}
    \toprule
    Stage & Elevated moderators & Dominant harm vector & Detection requirement \\
    \midrule
    Early childhood (0--7) & M1, M3 & Misattributed agency; unsafe instruction acceptance; displacement of adult-mediated learning & Elicited beliefs on system nature; instruction-following under unsafe prompts; substitution metrics \\
    \addlinespace[2pt]
    Middle childhood (8--12) & M1, M3, M5 & Cognitive offloading during foundational learning; algorithmic amplification of comparison content & Offloading events indexed to learning stage; exposure composition; comparison-content escalation \\
    \addlinespace[2pt]
    Adolescence (13--17) & M2, M3, M5 & Engagement optimization against reward sensitivity; parasocial dependency; crisis mishandling & Disengagement latency; AI-to-human interaction ratio; behavior under distress; escalation reliability \\
    \addlinespace[2pt]
    Emerging adulthood (18--25) & M3, M4 & Displacement of professional judgment; epistemic dependency prior to independent competence & Verification behavior under model uncertainty; trajectory of unaided task performance \\
    \addlinespace[2pt]
    Established adulthood (26--64) & M4 & Automation bias at maximum decision authority; hallucination harm multiplied by authorization scope & Override/challenge rates on erroneous outputs; confidence calibration; unaided performance tracking \\
    \addlinespace[2pt]
    Later adulthood (65+) & M1, M2, M4, M5 & Declining attribution capacity with high decision authority; synthetic-media susceptibility; social substitution & Calibration drift; response to persuasive/authority prompts; substitution ratios; capacity-sensitive consent \\
    \bottomrule
  \end{tabularx}
\end{table}

Two distribution features carry major governance implications:

\paragraph{Non-monotonic risk distribution.} Four of five moderators are elevated in later adulthood --- more than any other stage. However, current governance lacks elder-specific protective instruments, high-risk classifications, or test batteries calibrated to declining attribution capacity.

\paragraph{Inverse correlation of authority and scrutiny.} Established adulthood carries the highest decision authority yet receives the least developmental scrutiny. Its dominant harms stem from automation bias and authorization scope rather than individually unsafe outputs, rendering them invisible to content-based safety rubrics.

\subsection{Toward Lifespan Design Principles}

Substantive protective obligations already exist in binding and quasi-binding form, but almost exclusively for children. The UK Age Appropriate Design Code \citep{ico2020}, IEEE 2089-2021 \citep{ieee2089}, General Comment No.~25 to the Convention on the Rights of the Child \citep{crc2021}, and UNICEF's guidance on AI and children \citep{unicef2025} converge on a common set of commitments: design calibrated to developmental state, non-exploitation of immature reward systems, protection of identity formation, and meaningful supported control.

Our claim is that these commitments are not properly child-specific. They are lifespan principles that were codified first for children because that is where the evidentiary case was clearest and the political constituency strongest. Each maps onto a moderator that varies continuously across the lifespan rather than switching off at a statutory age boundary.

We therefore propose the Developmental Harm (DH) principles set out in Table~\ref{tab:principles}: a generalization of these instruments into stage-agnostic, moderator-indexed obligations. Three properties are worth stating explicitly.

\emph{The set is a proposal, not a codification.} We are not restating an existing enumerated standard. Table~\ref{tab:principles} is our synthesis, offered as a starting point for standards development rather than as a finished instrument.

\emph{The set is deliberately non-exhaustive and open to extension.} Each principle is anchored to a moderator or to a governance criterion we can currently defend. Moderators we have not identified, or harm classes that emerge from agentic deployment (Section~\ref{sec:future}), should generate additional principles. We expect DH9 onward to be written by others, and we regard extensibility as a design requirement rather than an admission of incompleteness.

\emph{The set requires no new statutory categories.} Because each principle generalizes an obligation already enacted somewhere for children, adoption is a question of extending scope, not of establishing an unprecedented duty --- a materially lower regulatory bar.

DH4 warrants comment: it has no child-specific antecedent, precisely because consequence authority is minimal in childhood. It is visible only once the matrix is read across the full lifespan --- which is the argument for constructing the matrix in the first place.

\begin{table}[t]
  \caption{Proposed Developmental Harm (DH) principles. Each generalizes an existing child-protective obligation into a lifespan form indexed to a moderator or governance criterion. The set is initial and extensible.}
  \label{tab:principles}
  \centering
  \small
  \begin{tabularx}{\linewidth}{@{}>{\raggedright\arraybackslash}p{0.9cm}L>{\raggedright\arraybackslash}p{5.6cm}@{}}
    \toprule
    & Proposed lifespan principle & Basis and index \\
    \midrule
    DH1 & Design calibrated to developmental state across the entire lifespan & Generalizes age-appropriate design \citep{ieee2089, ico2020}; indexes all moderators \\
    DH2 & Do not exploit affective or reward systems where self-regulation is reduced & Generalizes neurodevelopmental protection \citep{crc2021}; indexes M2 \\
    DH3 & Do not offload capabilities that the user is actively acquiring or maintaining & Extends educational protection duties \citep{unicef2025}; indexes M3 \\
    DH4 & Scale verification friction to the consequence authority being exercised & No child-specific antecedent; indexes M4 \\
    DH5 & Protect identity and social cohesion where substitution risk is elevated & Generalizes identity-formation protection \citep{crc2021}; indexes M5 \\
    DH6 & Verify supported decision-making where attribution capacity is reduced & Generalizes caregiver control \citep{ico2020, crc2021}; indexes M1 \\
    DH7 & Provide third-party verifiable safety attestation & Governance criterion; motivated by liability for deployed outputs \citep{moffatt2024} \\
    DH8 & Mitigate continuously post-deployment rather than at episodic pre-deployment checkpoints & Governance criterion; see Section~\ref{sec:tier3} \\
    DH9+ & (open) & To be specified as moderators and agentic harm classes are identified \\
    \bottomrule
  \end{tabularx}
\end{table}

\section{Why Internal Evaluation Cannot Close the Gap}
\label{sec:internal}

Self-evaluation by developer organizations is structurally incapable of ensuring safety. Commercial incentives, evaluator bias, and episodic testing create systematic blind spots that internal programs cannot overcome. This section diagnoses these three structural failures and examines precedents from high-stakes industries --- aviation, pharmaceuticals, and financial auditing --- that demonstrate why independent external oversight is mandatory.

\subsection{Three Structural Failures}

Systemic safety deficits stem directly from organizational and incentive architectures rather than individual incompetence.

\paragraph{Incentive misalignment.} Organizations evaluating safety hold commercial incentives to deploy. Evaluations are scoped to confirm launch readiness rather than maximize failure discovery, compressing adversarial timelines under competitive pressure \citep{bengio2025, bengio2026}. Developmental harms are deferred most under this architecture because they surface slowly and attribute poorly --- a dynamic recurring in automotive, pharmaceutical, and financial sectors.

\paragraph{The familiarity blind spot.} Internal red teams, composed predominantly of adult technical professionals, test for failures conceivable from their own developmental position. However, an adversarial prompt recognized as manipulation by a security researcher may appear as benign conversation to a child, a negotiable boundary to an adolescent, or incomprehensible to an impaired older adult \citep{apa2025}. Internal teams operating under deployment pressure lack structural motivation to integrate clinical, developmental, or geriatric expertise.

\paragraph{The coverage gap.} Pre-deployment testing is episodic, whereas deployed models interact with evolving populations, environments, and adversarial tactics. CrowdStrike reports that adversaries injected malicious prompts into production generative AI tools at more than ninety organizations during 2025, and that 42\% of exploited vulnerabilities were weaponized before public disclosure \citep{crowdstrike2026}: both are post-deployment phenomena invisible to pre-release testing. Monitoring practice has not kept pace. Fewer than half of organizations actively monitor production AI systems for accuracy, drift, and misuse, falling to 9\% among small firms \citep{gradientflow2025}; in the more heavily regulated medical device context, only 9\% of FDA-registered AI-based tools include a post-deployment surveillance plan \citep{dolin2026}. For developmental harms --- parasocial dependency, skill displacement, calibration drift --- harm accumulates across months of interaction and is undetectable in principle via static pre-release content rubrics.

\subsection{Cross-Industry Governance Precedent}

Mandating independent external oversight for safety-critical systems subject to commercial deployment pressure is standard practice across high-risk sectors:
\begin{itemize}[itemsep=2pt]
  \item Aviation rejects manufacturer self-attestation, requiring independent type certification and continuing airworthiness management across the operational lifecycle \citep{icao2018}.
  \item Pharmaceutical post-market surveillance mandates ongoing pharmacovigilance to track long-term adverse effects across diverse populations that pre-approval trials miss \citep{ema2017} --- the closest analogue to developmental harm tracking.
  \item Financial auditing mandates independent attestation because internal reporting incentives conflict with market transparency needs \citep{power1997, sox2002}.
\end{itemize}

Anthropic's Project Glasswing \citep{glasswing} demonstrates the feasibility of continuous external adversarial evaluation and coordinated disclosure. However, its published scope remains limited to a small number of security-focused organizations \citep{glasswing}, excluding clinical and developmental expertise. It establishes proof-of-concept for external oversight, but not broad ecosystem coverage.

\section{A Three-Tier Framework, Indexed by Developmental Stage}
\label{sec:framework}

This section specifies an external evaluation architecture designed to close the developmental safety gap. We establish core design criteria and an indexing principle that anchors evaluation to the Lifespan Vulnerability Matrix rather than testing methods alone. We then detail a three-tier operational framework --- automated simulation, expert red teaming, and post-deployment monitoring --- followed by structured escalation pathways and independence safeguards.

\subsection{Design Criteria and the Indexing Principle}

The evaluation framework satisfies five design criteria: institutional independence, operational continuity, breadth and depth, developmental specificity, and structured escalation.

Prior work has proposed layered technical and procedural designs for auditing large language models \citep{mokander2023} and end-to-end frameworks for internal algorithmic audit \citep{raji2020}. Our contribution is orthogonal to both: we specify what such audits must cover, indexed to the developmental state of the user, and we place the audit function outside the deploying organization.

Crucially, evaluation instruments are indexed to matrix risk cells rather than organized by method alone. Method-based tiering allows regimes to execute tests while leaving entire developmental profiles unassessed. Under our model, coverage is defined as the proportion of matrix cells with an assigned instrument, a defined baseline, and an escalation threshold --- providing an auditable metric usable for regulatory oversight, procurement, or underwriting.

\subsection{Tier 1: Automated Adversarial Simulation}

Synthetic agent populations continuously probe deployed systems across an adversarial matrix. Personas are specified by moderator profiles (e.g., low attribution capacity coupled with high consequence authority) rather than raw demographic tags. By including child, adolescent, and older-adult profiles as primary test cases, Tier 1 provides population-scale statistical coverage to detect low-frequency failure modes and generate performance baselines for Tier 3 tracking.

\subsection{Tier 2: Expert Human Red Teaming}

Human evaluation addresses nuanced harms outside predefined algorithmic rubrics. To resolve the familiarity blind spot, red team panel composition is directly informed by matrix requirements: child and developmental psychologists (Stages 1--3), human-factors specialists (Stages 4--5), and geriatric cognition experts (Stage 6), alongside offensive security analysts across all stages. Operating via triage, Tier 2 experts investigate and contextualize candidate signals flagged by Tier 1.

\subsection{Tier 3: Continuous Post-Deployment Monitoring}
\label{sec:tier3}

Adapting the pharmacovigilance model \citep{ema2017}, Tier 3 monitors production behavior using developmental cohort-stratified baselines. Aggregate baselines obscure population-specific anomalies (e.g., session-length spikes indicating healthy engagement in adults vs.\ parasocial dependency in adolescents). Continuous monitoring is uniquely capable of observing accumulating longitudinal harms --- including dependency, social substitution, skill displacement, and calibration drift.

\subsection{Structured Escalation}

Findings escalate based on moderator-weighted severity rather than raw incident counts. A single high-severity incident (e.g., low attribution capacity paired with high consequence authority) triggers escalation where high volumes of low-moderator events would not. Findings route to deploying organizations, sector regulators, and national cybersecurity bodies, bridging the gap between monitoring and regulatory enforcement.

\subsection{Institutional Form and Independence Safeguards}

While compatible with various structures --- such as independent standards nonprofits, accredited auditing firms, or government evaluation bodies --- the framework strictly mandates financial and organizational independence. Evaluators financially dependent on deploying entities replicate internal conflicts of interest. Credibility requires formal accreditation, conflict prohibitions, and enforceable reporting mandates established prior to deployment.

\section{Governance Recommendations}
\label{sec:governance}

Translating the Lifespan Vulnerability Matrix, the DH principles, and the three-tier framework into policy requires action across the AI governance ecosystem. This section outlines concrete recommendations for regulatory bodies, standards organizations, frontier AI developers, and institutional investors.

\paragraph{Regulatory bodies} --- including the EU AI Office, whose high-risk system obligations take effect in August 2026 \citep{euaiact}, the US AI Safety Institute, CISA, and sector regulators --- should replace binary age gating with developmental risk classification, classifying elder- and impairment-facing AI as high-risk by default. Incident reporting thresholds must adopt epidemiological models that aggregate low-severity signals to detect accumulating developmental harms before catastrophic events occur. Frontier models with autonomous exploitation capabilities warrant mandatory coordinated disclosure (analogous to the CVE process), while accreditation standards for external evaluators must mandate clinical and developmental expertise.

\paragraph{Standards bodies} (e.g., ISO, NIST, IEEE) must develop validated, age-stratified test instruments for the constructs in Table~\ref{tab:matrix}, such as attribution capacity, dependency, and skill displacement. The DH principles proposed in Table~\ref{tab:principles} are offered as a starting point for that work and are explicitly open to extension; IEEE 2089 \citep{ieee2089} provides an existing procedural template whose scope could be widened beyond children. Extending frameworks like the NIST AI RMF \citep{nist2023} into operational, developmentally stratified test protocols represents the highest-leverage standards contribution available.

\paragraph{AI developers} should adopt matrix coverage as an operational deployment requirement rather than a compliance posture, implementing the DH principles through external testing rather than internal checklists. Initiatives like Project Glasswing should be expanded in scope and multidisciplinary composition.

\paragraph{Boards, investors, and insurers} should require audited evidence of independent developmental coverage as a core governance control. Emerging litigation regarding addictive design \citep{metaag2023}, corporate liability for AI outputs \citep{moffatt2024}, and automated care denials \citep{lokken2023, psi2024} demonstrates that lifespan AI liability is a material risk. Matrix coverage provides an auditable, comparable metric directly usable for ESG disclosures, due diligence, and insurance underwriting.

\section{Discussion and Limitations}
\label{sec:limitations}

This section outlines methodological constraints and scope boundaries of the proposed framework. While independent evaluation shifts the failure distribution toward lower-severity events, it does not eliminate risk. Below, we address central limitations across causal attribution, construct validity, evaluator incentives, alignment scope, and empirical selection bias.

\paragraph{Causal attribution and cohort confounding.} Isolating AI-attributable harm from broader environmental, socio-economic, and educational influences remains a methodological challenge. Furthermore, longitudinal monitoring risks confounding developmental age effects with generational cohort effects (e.g., varying lifetime technology exposure across generations), requiring designs that systematically decouple them.

\paragraph{Construct validity.} The five moderators and six stages are grounded in developmental science but represent hypothesis structuring instruments rather than fully validated parameterizations. Their factor structure, independence, and measurement validity in human-AI interaction require ongoing empirical testing. The same caution applies to the DH principles: they are a proposed synthesis, and the mapping from principle to moderator is an argument we advance rather than a validated correspondence.

\paragraph{Evaluator incentive risks.} External evaluation bodies remain vulnerable to commercial capture. Mitigating conflicts of interest requires institutional safeguards adapted from financial auditing, such as mandatory evaluator rotation, whistleblower protections, and independent quality oversight \citep{sox2002}. We note the countervailing caution that formal audit regimes can decay into ritualized verification that produces assurance documents without producing assurance \citep{power1997}; an evaluation regime indexed to auditable coverage metrics is intended in part to resist that failure mode.

\paragraph{Scope boundaries.} The framework evaluates safety at the behavioral and output level; it serves as a necessary complement to, rather than a substitute for, foundational alignment research.

\paragraph{Empirical reporting biases.} The incident record exhibits strong selection bias --- overrepresenting high-severity cases from large organizations in high-income, Western contexts. Stages with sparse incident documentation (e.g., emerging adulthood) reflect an absence of systematic monitoring rather than an absence of risk. The global distribution of AI harm is broader and more concentrated in vulnerable populations than currently documented.

\section{Future Work}
\label{sec:future}

Future research must validate the developmental evaluation framework, formalize measurement standards, and address emerging technical challenges. We identify four priority avenues for empirical, methodological, policy, and cross-cultural investigation.

\paragraph{Empirical validation.} Research must verify whether each tier reliably surfaces distinct failure classes and quantify detection rate differentials between internal-only and hybrid external evaluation regimes. The M3 displacement hypothesis requires direct test in populations acquiring the offloaded capability, rather than extrapolation from adult offloading studies.

\paragraph{Measurement standardization.} Interdisciplinary efforts across clinical psychology, developmental science, gerontology, and HCI are needed to establish validated operational metrics for the five moderators, and to extend the DH principles beyond the initial set proposed here.

\paragraph{Political economy of certification.} Investigating the market and regulatory conditions --- such as liability law, procurement mandates, insurance underwriting, and network dynamics --- that drive adoption of independent safety evaluation.

\paragraph{Cross-cultural generalizability and agentic AI.} Testing the validity of the matrix's developmental constructs across non-Western populations and diverse regulatory jurisdictions. Crucially, extending the framework to autonomous agentic systems represents the most urgent technical challenge; agentic deployment fundamentally transfers consequence authority (M4) from the user directly to the system, and is the most likely source of DH principles beyond those proposed here.

\section{Conclusion}
\label{sec:conclusion}

This paper has argued that safety deficits in frontier AI are structural rather than procedural. Closing the assurance gap requires shifting from developer self-assessment to independent, external evaluation indexed directly to human developmental vulnerability.

The empirical record demonstrates that isolated product harms across the lifespan reflect a single underlying failure: systems are evaluated against an implicit model of a cognitively typical adult, ignoring stage-varying human vulnerability. Consequently, vulnerability peaks at both ends of the lifespan while oversight remains asymmetrical, leaving high-authority adult decisions largely unscrutinized and vulnerable populations exposed.

The operational tools --- automated simulation, expert red teaming, and continuous post-deployment monitoring --- already exist in prototype. What remains missing is their systematic integration into an independent governance architecture indexed to the developmental state of the user. Independent external evaluation is a structural necessity; the imperative now is how rapidly institutions can build the capacity to execute it.

\begin{ack}
The authors were members of OpenAI's External Red Teaming Network and want to thank OpenAI for foundational insights into adversarial evaluation methodology. The authors have no financial interests to disclose that bear on the governance arguments presented here.
\end{ack}

\section*{Ethics and Privacy Statement}

This paper discusses documented, publicly reported AI harm incidents, including cases involving suicide, child safety, CSAM, and elder exploitation, solely for the purpose of empirical governance analysis; no new or previously undisclosed harmful content is introduced. Individuals in cited incidents are identified only where their involvement is already a matter of public legal or journalistic record, and minors are not named. Allegations in pending or concluded litigation are described as allegations. The authors acknowledge that the empirical record carries documented selection and reporting biases, discussed in Section~\ref{sec:limitations}, and that any governance framework derived from it should be applied with attention to populations underrepresented in that record.

\end{document}